\documentclass[global,twocolumn]{svjour}
\usepackage{graphicx}
\usepackage{subfigure}

\begin{document}

\title{High contrast D$_{1}$ line electromagnetically induced transparency
in nanometric-thin rubidium vapor cell}
\author{Armen Sargsyan\inst{1} \and Claude Leroy\inst{2} \and Yevgenya Pashayan-Leroy\inst{2}
\and Rafayel Mirzoyan\inst{1} \and Aram Papoyan\inst{1} \and David Sarkisyan\inst{1} }
\offprints{A. Papoyan}
\mail{papoyan@ipr.sci.am, Institute for Physical Research, NAS of Armenia,
Ashtarak-2, 0203, Armenia}
\institute{Institute for Physical Research, NAS of Armenia, Ashtarak-2, 0203, Armenia
\and Laboratoire Interdisciplinaire Carnot de Bourgogne, UMR CNRS 5209 -
Universit\'{e} de Bourgogne, F-21078 Cedex, Dijon, France}
\maketitle

\begin{abstract}
Electromagnetically induced transparency
(EIT) on atomic D$_{1}$ line
of rubidium is studied
using a nanometric-thin cell with
atomic vapor column
length in the range of ${L=}$ \ 400 - 800 nm. It is shown that the
reduction of the cell thickness by 4 orders as compared with an ordinary
cm-size cell still allows to form an EIT resonance for $L= \lambda $
($\lambda =794$ nm) with the contrast of up to 40 $\%$. Remarkable
distinctions of EIT formation in nanometric-thin and ordinary cells are
demonstrated. Despite the Dicke effect of strong spectral narrowing and
increase of the absorption for $L=$ $\lambda /2$, EIT resonance is
observed both in the absorption and the fluorescence spectra for relatively
low intensity of the coupling laser. Well resolved splitting of the EIT
resonance in moderate magnetic field for $L=$ $\lambda $ can be used for
magnetometry with nanometric spatial resolution. The presented theoretical
model well describes the observed results.
\end{abstract}


%


%
%

\section{Introduction}
\label{intro} Continuous interest to
coherent population trapping (CPT) and the related
electromagnetically induced transparency
phenomena is stipulated
by a number of important applications in a variety
of fields such as laser
cooling, information storage, magnetometry, spectroscopy,
atomic frequency
references etc.
\cite{alzetta,arimondo,wynands,Fleisch,vanier,adams}.
EIT resonance can occur in a $\Lambda $-system with two
long-lived states and
one excited state coupled by two laser fields and
displays a strong
reduction in absorption where a maximum is expected
in the absence of the
coupling laser field (EIT can occur also in ladder
$\Xi $- and $V$-systems).
For many applications, it is important to reduce the
dimensions of a cell
containing atomic metal vapor where the EIT resonance
is formed, at the same
time keeping good resonance parameters. The linewidth
of the EIT resonance
in a $\Lambda $-system for the case of low laser
intensity is given by
$\gamma _{EIT}=2\Gamma _{21}+\Omega ^{2}/\gamma _{N}$ \cite{wynands}, where
$\Gamma _{21}$ is the coherence dephasing rate, with $\Omega $ being the Rabi
frequency and $\gamma _{N}$ the natural linewidth. For the case when
nanometric-thin cells (NTC) are employed \cite{sarkOptCom2001,sarkAPB2003,sarkPRA2006,josab},
the value of
$\Gamma_{21}$ is highly affected by atom-wall (cell window) collisions; it takes
place, though less pronounced, also for sub-millimeter thin cells \cite{knappe,fukuda}.
It is known that a unique collision with a dielectric
surface of an uncoated vapor cell is sufficient to thermalize the ground
hyperfine levels, with depolarization probability $0.5-1$ (see \cite{defreit}
and references therein). In order to preserve the atomic coherence in wall
collisions and to detect ultra-narrow linewidths with the help of coherent
processes like EIT, either coated walls or admixture of a buffer gas were
used \cite{knappe}. As the size of a vapor cell is reduced, the lifetime of
the ground-state coherence becomes shorter because of collisions of atoms
with cell windows: $\Gamma _{21}=(2\pi t)^{-1}$, where $t=L/u$ ($L$ is the
distance between windows and $u$ is the most probable thermal velocity).
Also the contrast of EIT resonance depends strongly on $\Gamma _{21}$.
Hence, one could expect that the EIT effect would completely wash out in the
case of $L<1$ $\mu $m ($2\Gamma _{21}>100$ MHz). Nevertheless, it was
demonstrated that the EIT resonance can be observed in thin cells with
thickness as small as $\sim $ 1 $\mu $m \cite{sarkPRA2006,josab}. The
explanation for this non-intuitive behavior is as follows: when the coupling
laser is resonant with an atomic transition, only the atoms flying nearly
parallel to cell windows and hence not experiencing wall collisions do
contribute to the formation of the EIT resonance.\\
\indent In \cite{stahler} it was shown that for an ordinary cm-size cell,
the excitation of the D$_{1}$ line results in greater EIT contrast as
compared to that for the D$_{2}$ line. One of the goals of the present study
was to verify whether this statement holds also for the case of NTC by
comparing the measured EIT contrast with that observed earlier for the
D$_{2} $ line~\cite{sarkPRA2006,josab}. A high contrast is needed, in
particular, to detect the splitting of the EIT resonance in an external
magnetic field for the thickness as small as $L=\lambda =$ 794 nm with a
better spectral resolution than that presented in \cite{SarkAPL2008}. This
may allow one to develop an optical magnetometer with nanometric-range
spatial resolution. Switching from D$_{2}$ line~to D$_{1}$ line~is also
expected to be favorable for forming the EIT resonance both in the
absorption and the fluorescence spectra when the cell thickness is
$\lambda/2$ (397 nm).

\section{Experiment}
\label{sec:Exp}

\subsection{Nanometric-thin cell}
\label{sec:NTC} 
\begin{figure}[h]
\centering\resizebox{0.3\textwidth}{!}{\includegraphics {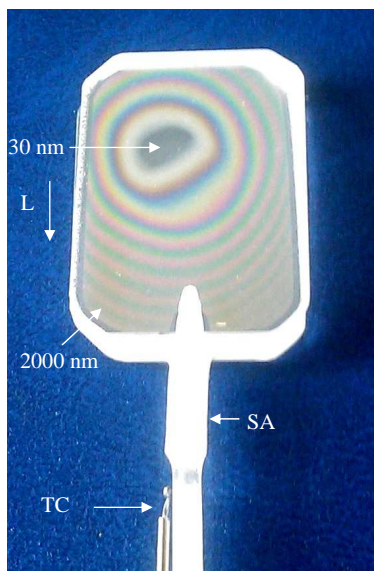}
}
\caption{Photograph of the NTC. The gap thickness $L$ varies from 30 to 2000
nm across the window as visualized by an interference pattern. Regions of $L=
$ 30 nm and 2000 nm are marked. SA: sapphire side-arm; TC: thermocouple. }
\label{fig:NTC}
\end{figure}

The design of a NTC is similar to that of extremely thin
cell described earlier \cite{sarkOptCom2001}. The modification implemented
in the present work is as follows. The rectangular 20 x 30 mm, 2.5 mm-thick
window wafers polished to $<$ 5 nm surface roughness are fabricated from
commercial optical grade sapphire (Al$_{2}$O$_{3}$), which is chemically
resistant to hot vapors of alkali metals. The wafers are cut across the $c$
-axis to minimize the birefringence. In order to exploit variable vapor
column thickness in the range of 30 - 2000 nm, the cell is vertically wedged
by placing a 2 $\mu $m-thick platinum spacer strip between the windows at
the bottom side prior to gluing. The NTC is filled with a natural rubidium
(72.2 $\%$ $^{85}$Rb and 27.8 $\%$ $^{87}$Rb). The photograph of the NTC
cell is presented in Fig. \ref{fig:NTC}.
Since the gap between the windows (i.e. the thickness of Rb atomic vapor
column) $L$ is of the order of visible light wavelength, an interference
pattern is clearly seen visualizing a smooth thickness variation from 30 nm
to 2000 nm. A thermocouple TC is attached to the sapphire side arm (SA) at
the boundary of metallic Rb to measure the temperature, which determines the
vapor pressure. The SA temperature in present experiment was 120 - 140~$^{\circ }$C,
while the windows temperature was kept some 20~$^{\circ }$C
higher to prevent condensation. This temperature regime corresponds to
atomic number density $N=$ 1.6 $\times $ 10$^{13}$ - 4.8 $\times $ 10$^{13}$
cm$^{-3}$. The NTC operated with a special oven with four optical outlets: a
pair of in line ports for laser beam transmission and two orthogonal ports
to collect the side fluorescence. This geometry allows simultaneous
detection of transmission and fluorescence spectra. The oven and NTC
assembly was rigidly attached to a translation stage for smooth vertical
movement to adjust the needed vapor column thickness without variation of
thermal conditions. For more details see \cite{sarkAPB2003,epj}.

\subsection{Experimental setup}
\label{sec:SetUp}

\begin{figure}[h]
\resizebox{0.47\textwidth}{!}{\includegraphics{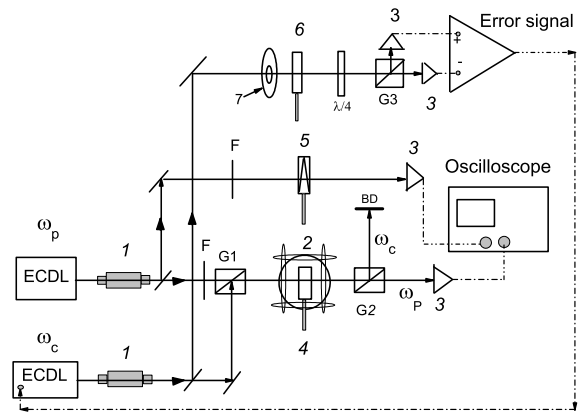}
}
\caption{Sketch of the experimental setup. \textit{1}- Faraday isolators;
G$_{1,2,3}$- Glan prisms; \textit{2}- Helmholtz coils; \textit{3}-
photodiodes; \textit{4}- main NTC; \textit{5}- auxiliary NTC of
$L=\lambda $; \textit{6}- auxiliary NTC of $L=\lambda /2$; \textit{7}-
permanent ring magnet (for details, see the text); BD- beam damp.}
\label{fig:SetUp}
\end{figure}
The experimental arrangement is sketched in Fig. \ref{fig:SetUp}. The beams
of two separate single-frequency extended cavity diode lasers (ECDL)
with $\lambda \approx $ 794 nm and $\sim $ 1 MHz
linewidth are well superimposed
and directed onto the NTC at near-normal incidence by the first Glan prism (G$_{1}$).
The coupling and probe beams have an elliptical shape of 3 $\times $
2 mm and have linear and perpendicular polarizations. Faraday isolators (\textit{1})
are used to prevent an extraneous feedback from reflected beams.
Nanometric-thin cell (\textit{4}) is placed inside the three pairs of
mutually perpendicular Helmholtz coils (\textit{2}) providing the
possibility to cancel laboratory magnetic field as well as to apply
homogeneous magnetic field. The transmission and fluorescence radiations are
recorded by the photodiodes (\textit{3}) followed by operation amplifiers;
the signal of the photodiodes is recorded by a two-channel digital storage
oscilloscope Tektronix TDS 3032B. The radiation power of the coupling and
probe lasers was varied throughout the measurement in the range of 1 - 30 mW
and 0.01 - 3 mW, respectively using neutral density filters \textit{F}. An
improved Dichroic-Atomic-Vapor-Laser-Locking\ technique realized with the
help of a separate NTC with $L=\lambda /2$, a permanent ring magnet (\textit{7}),
$\lambda /4$
plate, Glan prism (G$_{3}$), and an error signal unit for
electronic subtraction of two photodiodes signals is used for the coupling
laser frequency stabilization (see \cite{epj}). The frequency
reference spectrum was formed with the help of an auxiliary NTC (\textit{5})
with $L=\lambda $ \cite{epj}. A second Glan prism (G$_{2}$) was used
for separating the coupling and probe beams, so that only the probe beam
transmission could be monitored.

\subsection{EIT in nanometric-thin cell: the coupling laser is resonant with
atomic transition}
\label{sec:CoupleRes}
\begin{figure}[h]
\resizebox{0.47\textwidth}{!}{\includegraphics{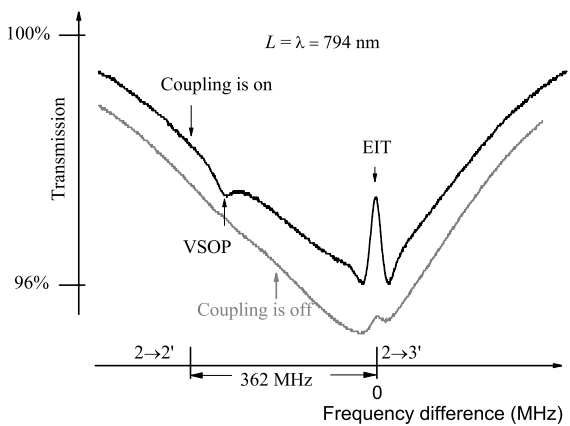}
}
\caption{EIT spectrum for $^{85}$Rb D$_{1}$ line (upper curve). $L=\lambda =$ 794 nm.
Coupling laser is resonant with $F=3\rightarrow F^{\prime
}=3$ transition, probe laser is scanned across $F=2\rightarrow F^{\prime
}=2,3$ transitions. Grey line: the spectrum without the coupling laser
(vertically shifted for convenience).}
\label{fig:Fig3}
\end{figure}

\begin{figure}[h]
\resizebox{0.47\textwidth}{!}{\includegraphics{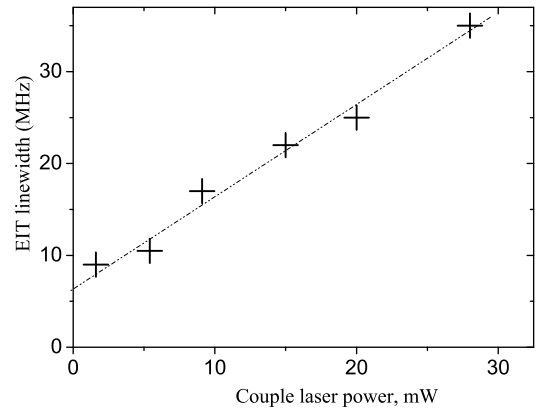}
}
\caption{Dependence of $^{85}$Rb D$_{1}$ line EIT resonance linewidth on
coupling laser power for $\Delta =0$. $L=\lambda =$ 794 nm.}
\label{fig:Fig4}
\end{figure}
\begin{figure}[th]
\label{fig:Fig5} \centering
\subfigure[]{
\resizebox{0.47\textwidth}{!}{\includegraphics{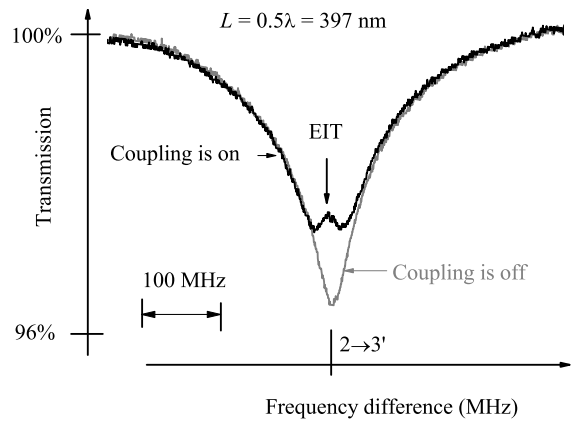}
}
\label{fig:5a}
} \hspace{0.5cm}
\subfigure[]{
\resizebox{0.47\textwidth}{!}{\includegraphics{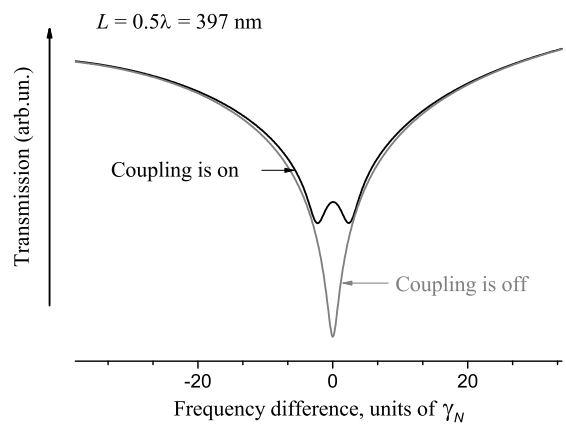}
}
\label{fig:5b}
}
\caption[Optional caption for list of figures]{Probe laser transmission
spectrum for $^{85}$Rb D$_{1}$ line. $L=\lambda /2=$ 397 nm.
Coupling laser ($P_{c}=$ 27 mW) is resonant with $F=3\rightarrow F^{\prime
}=3$ transition, probe laser ($P_{p}=$ 0.04 mW) is scanned across $F=2\rightarrow F^{\prime }=2,3$.
Grey line: spectrum when coupling laser is
off. Graph \textit{a}: experiment; graph \textit{b}: numerical calculations
for the experimental parameters, $\Omega_{c}=1.5\gamma_{N}$, $\Omega_{p}=0.06\gamma_{N}$,
$\gamma_{21}^{\prime }=$ 6 MHz. }
\end{figure}
\begin{figure}[th]
\centering
\subfigure[]{
\resizebox{0.47\textwidth}{!}{\includegraphics{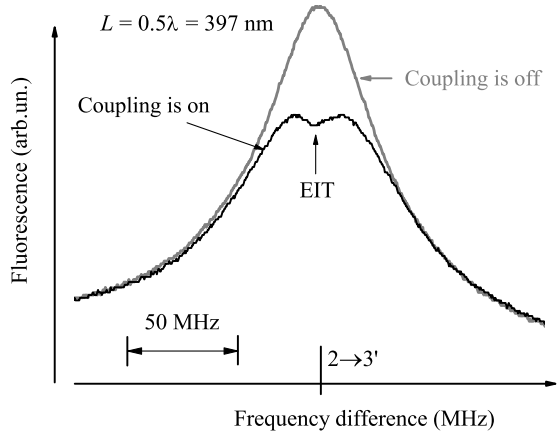}
}
\label{fig:6a}
} \hspace{0.5cm}
\subfigure[]{
\resizebox{0.47\textwidth}{!}{\includegraphics{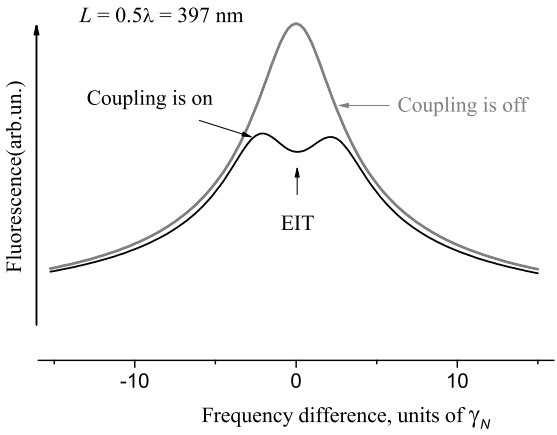}
}
\label{fig:6b}
}
\caption[Optional caption for list of figures]{Transmission spectrum of the
probe laser for $^{85}$Rb D$_{1}$ line, $L=\lambda /2=$ 397 nm.
Coupling laser ($P_{c}=$ 27 mW) is resonant with $F=3\rightarrow F^{\prime
}=3$ transition, probe laser ($P_{p}=$ 3 mW) is scanned across $F=2\rightarrow F^{\prime }=2,3$.
Grey line: the coupling laser is off. Graph
\textit{a}: experiment; graph \textit{b}: numerical calculations for the
experimental parameters, $\Omega _{c}=1.5\gamma _{N}$, $\Omega
_{p}=0.5\gamma _{N}$, $\gamma _{21}^{\prime }=$ 6 MHz. }
\label{fig:Fig6}
\end{figure}

In Fig. \ref{fig:Fig3} is shown the transmission spectrum of the probe laser
(the upper curve) for the thickness $L=\lambda $ ($\lambda $ = 794 nm) when
the 20 mW coupling laser radiation is resonant with $F=3\rightarrow
F^{\prime }=3$ transition, while the 0.04 mW probe laser radiation frequency
is scanned across the $F=2\rightarrow F^{\prime }=2,3$ transitions (see Fig.
\ref{fig:Levels} in case of zero coupling laser detuning $\Delta =0$). The
full width at half maximum (FWHM) of the EIT resonance is 25 MHz, while the
contrast defined as the ratio of the EIT amplitude to the height of the
shoulders is 34 $\%$. For a higher coupling laser radiation power, 28 mW,
the the linewidth broadens to $\sim $ 35 MHz, while the contrast reaches
nearly 40 $\%$. Note that the best contrast for D$_{2}$ line obtained under
similar experimental conditions was 5 - 6 $\%$ \cite{sarkPRA2006,josab}.
The grey line shows the transmission spectrum when the coupling laser is
off. The side-arm temperature of NTC is 140 $^{\circ }$C (Rb atomic vapor
density $N=$ 4.8 $\times $ 10$^{13}$ cm$^{-3}$). As is seen, the velocity
selective optical pumping (VSOP) peak demonstrates an increase of the
absorption, since the coupling laser transfers a number of atoms from $F=3$
to $F=2$, thus increasing the probe absorption for $F=2\rightarrow F^{\prime}=2,3$
transitions (also the EIT resonance is superimposed on the VSOP
at $F=2\rightarrow F^{\prime }=3$ transition).
The theoretical spectrum as
obtained from the numerical calculations (see Section 3) is shown in
Fig. \ref{fig:Fig7} (the second curve from the bottom, case of $\Delta =0$) for
the same parameters as in Fig. \ref{fig:Fig3} except the probe laser power
taken to be 1 mW. \\
\indent Fig. \ref{fig:Fig4} presents the EIT resonance linewidth as a
function of the coupling laser power for $\Delta =0$. The intersection of
the curve with the vertical axis gives the residual coherence dephasing rate
of $\gamma_{21}^{\prime }$ = 6 MHz, which is caused by several reasons,
including atom-atom, atom-wall collisions in NTC, time-of-flight broadening,
as well as due to the fact that the lasers are not coherently coupled \cite{wynands}.\\
\indent Earlier, it was demonstrated that there is a dramatic difference for
the absorption (and also fluorescence) spectrum when the thickness of NTC is
reduced from $L=\lambda $ to $L=\lambda /2$, which is caused by the revival
of Dicke-type coherent narrowing effect. Particularly, the linewidth of the
sub-Doppler absorption spectrum for $L=\lambda /2$ is $\sim $ 3 times
narrower than that for $L=\lambda $, while showing the same peak absorption
value \cite{Dutier2003,BlochPRA2004,CartalPRA2007}. Moreover, as it was
recently demonstrated, for $L=\lambda /2$ it is impossible to form any type
of sub-Doppler dip (VSOP resonance) neither in transmission nor in
fluorescence spectra, even for the laser intensity as high as several tens
of W/cm$^{2}$, as opposed to $L=\lambda $ case where only a few mW/cm$^{2}$
probe laser intensity is sufficient for a dip formation. For this reason,
the $L=\lambda /2$ case is of a particular interest for EIT resonance
formation.\\
\indent The striking point is that it is possible to form an EIT resonance
(i.e. a peak of reduced absorption) for the thickness $L=\lambda /2$ already
at a coupling laser intensity as low as $\sim $ 0.1 W/cm$^{2}$ (see Fig. \ref{fig:5a}).
The grey curve shows the 120 MHz-wide sub-Doppler transmission
spectrum for $L=\lambda /2$ when the coupling laser is off. The EIT for the
thickness $L=\lambda /2$ was theoretically addressed in \cite{josab}.\\
\indent The EIT resonance demonstrating reduction of the fluorescence (dip
of reduced fluorescence) for $L=\lambda /2$ is shown in Fig.~\ref{fig:6a}.
The grey curve shows the sub-Doppler fluorescence spectrum of 70 - 80 MHz
linewidth \cite{sarkAPB2003} when the coupling laser is off. The theoretical
spectra for the parameters matching exactly the corresponding experimental
conditions of Figs. \ref{fig:5a} and \ref{fig:6a} with $\Omega
_{c}=1.5\gamma _{N}$, $\gamma _{21}^{\prime }=$ 6 MHz are shown in Fig. \ref{fig:5b}
(transmission spectrum, $\Omega _{p}=0.06\gamma _{N}$,) and Fig. \ref{fig:6b}
(fluorescence spectrum, $\Omega _{p}=0.5\gamma _{N}$). As it is
seen, there is a good agreement between the experiment and the theory. Thus,
despite the Dicke effect of strong spectrum narrowing and increase of the
absorption for $L=\lambda /2$, the increase of coupling laser intensity to $\sim $
100 mW/cm$^{2}$ makes it possible to form an EIT resonance both in
the absorption and the fluorescence spectra.

\subsection{EIT in nanometric-thin cell: dependence on coupling laser
detuning}

\label{sec:CoupleDetun}

\begin{figure}[th]
\centering
\subfigure[]{
\resizebox{0.47\textwidth}{!}{\includegraphics{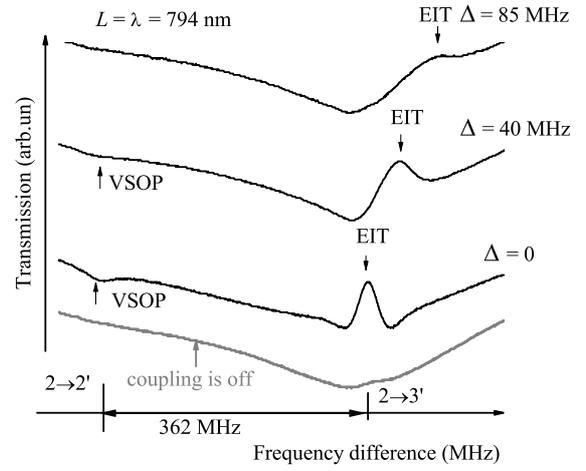}
}
\label{fig:7a}
} \hspace{0.5cm}
\subfigure[]{
\resizebox{0.47\textwidth}{!}{\includegraphics{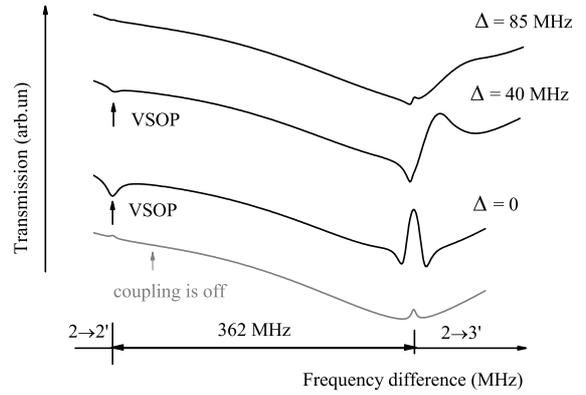}
}
\label{fig:7b}
}
\caption[Optional caption for list of figures]{EIT spectra for $L=\lambda =$ 794 nm
when the coupling laser is detuned by $\Delta $ from $ F=3\rightarrow F^{\prime }=3$ transition.
The spectra are shifted vertically
for convenience. The lower grey curve is the transmission spectrum when the
coupling laser is off. Graph \textit{a}: experiment; graph \textit{b}:
numerical calculations for the experimental conditions, $\Omega_{c}=1.3\gamma_{N}$,
$\Omega_{p}=0.3\gamma_{N}$, $\gamma_{21}^{\prime }=$ 6 MHz. Additional dephasing rate $\gamma_{21}^{\prime }$($\Delta $)
is responsible for the EIT behavior seen for $\Delta \neq 0$ (see the text).}
\label{fig:Fig7}
\end{figure}
\begin{figure}[h]
\includegraphics {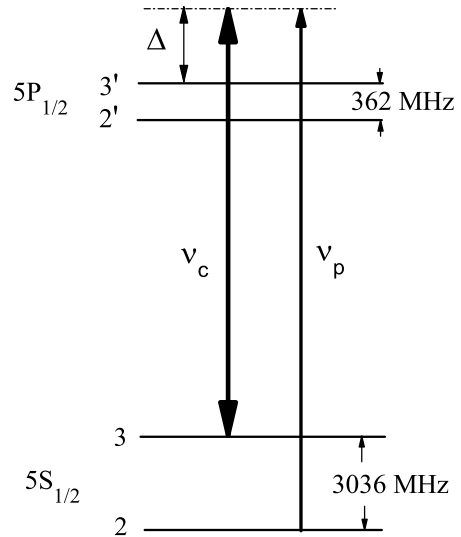}
\caption{Relevant energy levels of $^{85}$Rb D$_{1}$ line involved in EIT
process. The coupling laser detuning $\Delta $ varies from $F=3\rightarrow
F^{\prime }=3$ transition in the range of $0-100$ MHz.}
\label{fig:Levels}
\end{figure}

One of the drastic differences in behavior of EIT realized in NTC and in an
ordinary 1 - 10 cm long cell is exhibited by the EIT linewidth and amplitude
dependence on the coupling laser detuning $\Delta $ (see Fig. \ref{fig:Levels}).
The EIT linewidth and amplitude (contrast) as a function of
blue detuning $\Delta $ of the coupling laser for $L=\lambda $ can be
evaluated from the experimental and theoretical transmission spectra shown
in Fig. \ref{fig:Fig7}. The side-arm temperature of NTC is 140 $^{\circ }$C,
$P_{c}\approx $ 20 mW. In order to get more prominent EIT resonance for the
large values of detuning, the probe laser radiation power was increased to 1
mW. As it is seen, the increase of detuning $\Delta $ from 0 to 40 and 85
MHz results in rapid broadening of EIT resonance linewidth from 25 MHz to 50
and 80 MHz FWHM, correspondingly, meanwhile the EIT\ amplitude rapidly
decreases. This behavior has the following physical explanation. For the
case of an exact atomic resonance ($\Delta =0$), mainly the atoms flying
parallel to NTC windows contribute to formation of the EIT resonance. The
interaction time for these atoms is $\tau_{D}=D/v$, where $D$ is the laser
beam diameter ($D\gg L$). When the coupling laser is detuned from an atomic
transition by a value of $\Delta $, only those atoms having the velocity
projection $v_{z}=2\pi \Delta /k$ contribute to the formation of EIT, and
for these atoms the flight time between the cell windows shortens with the
detuning increase: $\tau_{L}=L/v_{z}=kL/(2\pi \Delta )$. This causes
frequent quenching collisions of atoms with the cell windows leading to the
increase of $\gamma_{21}^{\prime }$ and consequently, to strong linewidth
broadening and contrast reduction of the EIT resonance. For this case $\gamma_{EIT}$
can be described by a simple expression
\begin{equation}
\gamma_{EIT}(\Delta )=\gamma_{EIT}\big |_{\Delta =0}+2\gamma_{21}^{\prime
}(\Delta ),  \label{eq:Gamma_EIT}
\end{equation}%
where $\gamma_{21}^{\prime }(\Delta )=1/(2\pi \tau_{L})$. Replacing the
gap thickness by a justified parameter $L_{eff}=L/2$, we get
$\gamma_{21}^{\prime }(\Delta )=\Delta /\pi $
for $L=\lambda $. From (\ref{eq:Gamma_EIT})
for $\Delta =$ 40 and 85 MHz we have $\gamma_{EIT}=$ 50 and
79 MHz, correspondingly, which is in good agreement with the experimental
values. We should note that measurement of EIT resonance linewidth $\gamma_{EIT}$
at large coupling laser detuning could serve as a convenient tool to
study atom-window collisions and properties of window material.\\
\indent This behavior completely differs from that observable in ordinary 1
- 10 cm long cell where formation of EIT resonance on the wings of the
Doppler profile of the absorption line can lead to its narrowing \cite{Wang}
as opposed to the case of NTC.\\
\indent It is well known that addition of a buffer gas into 1 - 10~cm long
ordinary cell improves EIT resonance parameters \cite{wynands,Fleisch}. The
preliminary results have shown that admixture of 20 Torr of N$_{2}$ buffer
gas to Rb vapor in NTC leads to decrease of EIT broadening rate with the
increase of $\Delta $. This is explained by substantial shortening of the
mean free path of Rb atoms due to Rb-N$_{2}$ collisions ($\sim $ 300 nm),
which becomes less than the NTC thickness ($L=$ 794 nm). However, at $\Delta
=0$ there appears undesirable supplementary broadening: Rb atoms flying
parallel to the NTC windows experience velocity changing collisions with N$_{2}$,
change their direction and collide with the windows, which leads to
the increase of $\Gamma_{21}$.

\subsection{Splitting of EIT resonance in magnetic field, $L=\protect\lambda$}

\label{sec:SplitEIT}

\begin{figure}[h]
\resizebox{0.47\textwidth}{!}{\includegraphics {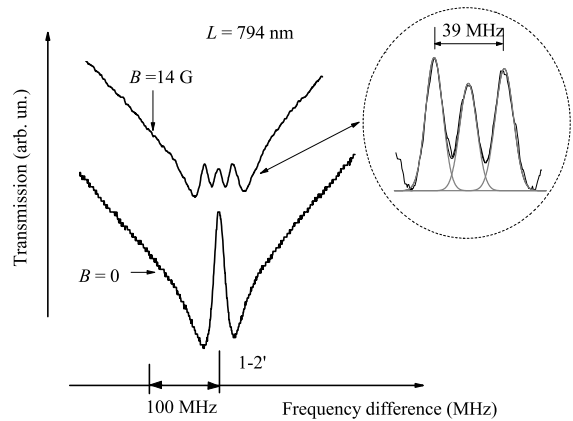}
}
\caption{Probe laser transmission spectrum for $L=\lambda =$ 794 nm
with a longitudinal magnetic field $B=$ 14 G (upper curve). $P_{c}=$ 5 mW,
$P_{p}=$ 0.05 mW. Splitting of EIT resonance to three components is well
seen. The inset shows the results of fitting by a Gaussian function (grey
lines); FWHM of an individual EIT component is 10 MHz. Lower curve: the EIT
resonance for $B=0$. }
\label{fig:Split}
\end{figure}

In \cite{sarkPRA2006} it was shown that using NTC with Rb vapor of
$L=2\lambda $ thickness ($\lambda =$ 780 nm), one can measure a magnetic
field by splitting EIT resonances formed at D$_{2}$ line excitation.
However, due to a small contrast of the EIT resonance it was impossible to
detect the splitting at smaller $L$ values. Switching to D$_{1}$ excitation
allowed us to improve the EIT contrast and thus to detect the splitting for
$L=\lambda =$ 794 nm as is shown in Fig. \ref{fig:Split}. The measurement
shown in the upper curve was done with application of $B=$ 14 G longitudinal
magnetic field ($\mathbf{B}\Vert \mathbf{k}$), for laser radiation powers
$P_{c}=5$ mW and $P_{p}=0.05$ mW. Three well resolved EIT resonances are
clearly seen in the upper curve; the lower curve presents the EIT resonance
when $B=0$. In \cite{sarkPRA2006} it was shown that in this case three
$\Lambda $-systems are formed, and the frequency separation between these EIT
components is 1.4 MHz/G $\times $ 14 G $=$ 19.6 MHz. The inset presents the
results of fitting by Gaussian function resulting in the linewidth of 10 MHz
FWHM for an individual EIT component. Note that the exact splitting value
for strong $B$ fields is given by the Rabi-Breit formula \cite{wynands}.\\
\indent So-called "$\lambda $-Zeeman technique" (LZT) to investigate atomic
transitions in magnetic field was presented in \cite{SarkAPL2008}. Rb atoms
are confined in a NTC with $L=\lambda $. Narrow VSOP resonances of 30 - 35
MHz linewidth formed in a single beam in transmission spectrum are split
into several components in a magnetic field; their frequency positions and
transition probabilities depend on the $B$ field. As one can see in Fig. \ref{fig:Split},
the observed splitting of EIT resonance provides a spectral
resolution, which is better than that for LZT by a factor of 3 \cite{SarkAPL2008}.
Moreover, among the inherent advantages of NTC is the
possibility to apply very high magnetic field using widely available strong
permanent ring magnets (our Helmholtz coils produce $B<$ 200 G). In spite of
a strong inhomogeneity of the $B$ field (it can reach $\sim $ 150 G/mm in
axial direction), the variation of $B$ inside the atomic vapor column does
not exceed $\sim $ 0.1 G (several orders less than the applied $B$ value)
because of sub-micrometer thickness of the NTC. Our preliminary results show
that it is realistic and straightforward to determine a spatially-localized
value of a $B$ field in the range of 1 - 1000 G with nearly the same
precision throughout the whole range, by measuring the frequency interval
between the EIT components. This is especially valuable for mapping of
strongly non-homogeneous magnetic field.

\section{Theoretical model}

\label{sec:Theory}

\begin{figure}[h]
\centering\includegraphics {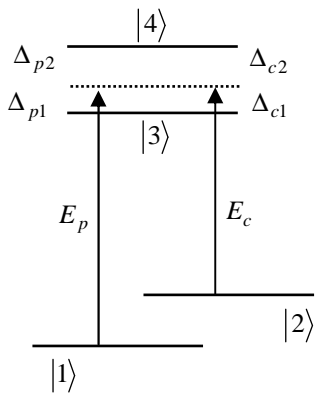}
\caption{$\Lambda $-type four-level atomic system.}
\label{fig:AtScheme}
\end{figure}

To calculate the observed EIT resonances we consider four-level $\Lambda $-type
atomic systems consisting of two ground hyperfine levels ($F_{g}=2,3$)
and two excited levels ($F_{e}=2,3$) in the D$_{1}$ line of $^{85}$Rb (see
Fig. \ref{fig:Levels}) confined in a cell of thickness $L$. The probe field
$E_{p}$ (with center angular frequency $\omega_{p}$) and the coupling field
$E_{c}$ (with angular frequency $\omega_{c}$) couple the transitions
$|1\rangle \rightarrow |3\rangle ,|4\rangle $ and $|2\rangle \rightarrow
|3\rangle ,|4\rangle $, respectively (see Fig. \ref{fig:AtScheme}). For
simplicity, we restrict ourselves to one dimensional situation where the two
driving linearly polarized laser radiations are in the $\pm z$-direction.
The probe and coupling lasers are detuned from the corresponding atomic
resonances by
\begin{equation}
\begin{array}{ll}
\label{eq:Detunings}\Delta_{p1,p2}=\omega_{p}-\omega_{3,4;1}-k_{p}v_{z},
&  \\
\Delta_{c1,c2}=\omega_{c}-\omega _{3,4;2}-k_{c}v_{z}, &
\end{array}
\end{equation}
where the subscripts $p(c$) mark the detunings for the probe (coupling)
lasers, $\omega_{ij}$ is the frequency difference between levels $i$ and $j$,
and $k_{p,c}v_{z}$ are the Doppler shifts for atoms corresponding to the
velocity component $v_{z}$ along the propagation vectors $\mathbf{k}_{p,c}$.
The total Hamiltonian of the system is given by $\hat{H}=\hat{H}_{0}+\hat{H}_{I}$,
where $\hat{H}_{0}$ describes a free atom and $\hat{H}_{I}$ describes
the interaction between the atom and the electric fields. With
electric-dipole and rotating-wave approximations, the interaction
Hamiltonian $\hat{H}_{I}$ of the system is given by
\begin{equation}
\begin{array}{ll}
\hat{H}_{I}= & -\hbar \big (\Omega _{p1}|3\rangle \langle 1|+\Omega_{p2}|4\rangle \langle 1| \\
& +\Omega_{c1}|3\rangle \langle 2|+\Omega_{c2}|4\rangle \langle 2|\big)+\mathrm{H.c.},
\end{array}
\label{eq:IntHamilt}
\end{equation}%
where $\Omega _{p1,p2}=\mu _{1;3,4}E_{p}/2\hbar $ and $\Omega _{c1,c2}=\mu
_{2;3,4}E_{c}/2\hbar $ are the Rabi frequencies of the probe and coupling
fields, respectively, with $\mu _{ij}$ being the electric-dipole matrix
element associated with the transition from $j$ to $i$, and H.c. represents
Hermitian conjugate.
To describe the dynamics of the system driven by this Hamiltonian we adopt
here the theoretical model presented in Ref. ~\cite{josab}. Interested
readers may refer to this work and to the references therein for the
mathematical details of the calculations. We summarize here the main
features of the approach used. The basic assumptions made in the model are
as follows: the atomic number density is assumed to be low enough so that
the effect of collisions between the atoms can be ignored; the atoms
experience inelastic collisions with the cell walls; the incident beam
diameters largely exceed the cell thickness. The internal atomic dynamics is
described by a semiclassical density matrix $\rho =\sum {\rho_{ij}|i\rangle
}\langle j|$, the time evolution of which is given by the Liouville equation
of motion :
\begin{equation}
\dot{\rho}=-i/\hbar \lbrack \hat{H},\rho ]+\Gamma \rho ,  \label{eq:Density}
\end{equation}%
where $\Gamma $ is the relaxation operator :
$\Gamma_{ij}=\frac{1}{2}(\gamma_{i}+\gamma_{j})+\gamma_{ij}\delta_{ij}$.
Here $\gamma_{i}$, $%
\gamma_{j}$ are the natural decay rates of states $|i\rangle $ and $|j\rangle $,
and $\gamma_{ij}$ is the pure dephasing rate. \newline
\indent To obtain the probe transmission spectra, we provide an exact
numerical solution of the set of optical Bloch equations deduced from Eq. (\ref{eq:Density})
and coupled to Maxwell equations~\cite{Shore}. In
accordance with the assumptions made, we neglect the collisional broadening
of optical transitions compared to the spontaneous decay rate, the
relaxation of atoms traveling out of the diameter of the laser beam. The
dephasing rate of the ground-state coherence, $\Gamma_{21}$, is determined
in the model by the finite time of flight of atoms between the cell windows.
This parameter, usually introduced into theoretical calculations as a
phenomenological constant to fit the experimental data, is taken into
account in the model exactly by solving the temporal density-matrix
equations with proper boundary conditions for each atom separately. \\
\indent A number of relevant parameters, both inherent to the atomic system
and the experimental setup, are included into the model such as Doppler
broadening, phase fluctuations in the laser fields~\cite{Agarwal,Knight},
non-correlation of the laser fields~\cite{wynands}. The light reflections on
the locally highly-parallel cell walls of a nano-cell behaving as a Fabry-P\'{e}rot
cavity \cite{FabryPerot} is taken into account. For the Rabi
frequency the estimates are obtained from $\Omega /2\pi =a\gamma (I/8)^{1/2}$,
where $I$ is the laser intensity in mW/cm$^{2}$, $\gamma $ is the decay
rate of the excited state, and $a$ is a fit parameter (for our case $a\approx 0.2$) \cite{Krmpot}.
A general agreement between the results of our
calculations and that of the experiment is seen.

\section{Discussion}

\label{sec:Discussion}

Miniaturization of EIT setup without loss of resonance characteristics is
among the challenges for technological implementation of the
electromagnetically induced transparency. The results obtained in the
present study are not intuitive, and it is worth to compare them with the
relevant results presented in literature, and analysing the distinctions.
In \cite{Wei} a similar setup was employed for the EIT studies: two separate
lasers of 1 MHz linewidth and approximately the same power were used to form
the EIT resonance on D$_{1}$ line of Rb. The vapor was confined in a 3
cm-long glass cell. The following key parameters have been achieved: $\sim $
60 $\%$ for the EIT resonance contrast, and $\sim $ 20 MHz for the spectral
width. The corresponding values obtained in the present work with the use of
$L=$ 794 nm NTC are $\sim $ 40 $\%$ and $\sim $ 35 MHz, respectively. Thus,
decreasing the cell length by a factor of $3.8\times 10^{4}$ does not lead
to a dramatic deterioration of the key parameters of EIT.\\
\indent Sub-millimeter-thick cells have been used in \cite{knappe,fukuda}
for EIT formation. In recent works \cite{Ghosh,Gaeta} EIT resonance is
formed in Rb vapor generated in hollow-core photonic band-gap (HCPBG) of
inner diameter 6 $\mu $m and $L=$ 25 cm. Large $L$ together with high laser
intensity achieved in HCPBG leads to the decrease of threshold laser power
for EIT formation. However, the width of EIT resonance reaches $\sim $ 100
MHz, which exceeds by an order of magnitude the width of EIT resonance in
NTC. Moreover, the realization of HCPBG method is rather complicated
technically.\\
\indent There are yet other papers where suggestions to improve the EIT
characteristics in thin cells are addressed. In \cite{Litvinov} formation of
narrow EIT resonance in nano-cells is proposed by using supplementary
anti-relaxation coating for windows, aiming at decrease of coherence
dephasing rate. However, the realisation of such a coating is not yet
feasible. Formation of regions of few tens of micrometer in quartz
substrate-window is realized using chemical etching with subsequent filling
these regions with Rb vapor \cite{baluk}, meanwhile the realization of EIT in
such a structure has not been yet discussed. Based on the above analysis, we
believe the most convenient and technically justified option for obtaining
narrow-linewidth and high-contrast EIT resonances is the NTC scheme
presented in this work (note that the NTC has been successfully used also
for EIT registration in a ladder $\Xi $-system \cite{LadderArm}).\\
\indent As it follows from the obtained results, the decrease of $L$ is
compensated by an increase of Rb atomic density, so that the product $N\times L$
practically does not change. An important particularity of EIT is
the weak influence of 5P$_{1/2}$ level broadening on the EIT parameters (in
contrast, 5P$_{1/2}$ level broadening dramatically affects the amplitude of
the VSOP resonance). The latter can allow one to form EIT resonance for $L<100$ nm
by increasing the density $N$ and improving the laser parameters
(for example, by the technique of phase-locking the coupling and probe
lasers \cite{Benson}), realization of lock-in detection, etc. Such a small
thickness of the NTC is of interest for the study of van der Waals
interaction of atoms with dielectric windows \cite{Fichet}.

\section{Conclusion}

\label{sec:Conclusion}

The peculiarities of the EIT phenomenon are studied in transmission and
fluorescence spectra under D$_{1}$ line excitation of rubidium atoms with
the help of nanometric-thin vapor cell with a possibility to exploit vapor
column thickness $L$ in the range of 400 - 800 nm. The striking point of the
present work is that the reduction of $L$ by more than 4 orders as compared
with an ordinary cm-size cell does not worsen the high contrast of EIT peak
(up to 40 $\%$). This relatively weak thickness dependence is explained by
involvement of atoms with particular velocity components in EIT resonance
formation: when the coupling laser is resonant with the atomic transition
only the atoms moving parallelly to the NTC windows contribute to the EIT
formation, and the quenching collisions with the windows are not essential.
The situation is dramatically different when the coupling laser is detuned
from the atomic transition. In this case the atoms having velocity component
towards the windows are responsible for the EIT, they experience collisions
with the windows resulting in broadening and contrast worsening of the
resonance. It is justified that with the help of the NTC with $L=\lambda $
and for a large frequency detuning $\Delta $ of the coupling laser from
atomic resonance it is possible to directly reveal and evaluate the strong
influence of atom-walls collisions on the EIT linewidth. By changing the
value of $\Delta $ one can smoothly control the EIT linewidth and contrast;
measurement of the EIT linewidth makes it is possible to determine the
atom-wall collision rate. The theoretical model well describes the observed
results. Formation of the EIT resonance in a $\Lambda $-system for even
smaller thickness (below 100 nm) can offer a possibility for quantitative
study of atom-surface van der Waals interaction.\\
\indent For the first time, it is demonstrated that despite the Dicke-effect
of strong spectrum narrowing and increase of the absorption when thickness
is $L=\lambda /2$ (397 nm), the relatively low intensity of the coupling
laser is sufficient to form the EIT resonance both in the absorption and the
fluorescence spectra. This is the minimal $L$ for the coherent process
reported so far.\\
\indent It is shown that the EIT resonance in a NTC with $L=\lambda $ splits
in an external magnetic field to well-resolved components that may allow one
to develop magnetometer with nanometric-range spatial resolution. This can
be of importance for measuring and mapping the strongly non-homogeneous
magnetic field.\newline

\section{Acknowledgement}

The authors are grateful to A. Sarkisyan for his valuable participation in
fabrication of the NTC and to Charles Adams for useful discussions. Research
conducted in the scope of the International Associated Laboratory IRMAS.
Armenian team thanks for the ANSEF Opt-2428 support. R. M. acknowledges the
support from the NFSAT, NAS RA and CRDF (grant no. ECSP-10-10 GRSP).


\end{document}